\documentclass{article}

\usepackage{spconf}
\usepackage{amssymb}
\usepackage{amsmath}
\usepackage{epsfig}

\title{The Bispectral Aliasing Test: A Clarification and Some Key Examples}
\name{Kevin R. Vixie, Murray Wolinsky and David E. Sigeti
\thanks{This work was supported by Los Alamos National Laboratory LDRD 97028.}}
\address{Los Alamos National Laboratory \\
                          Los Alamos, NM 87545 \\
                          murray@lanl.gov}
\begin{document}
\maketitle
\begin{abstract}
Controversy regarding the correctness of a test for aliasing
 proposed by Hinich and Wolinsky~\cite{hw} has been surprisingly
 long-lived. Two factors have prolonged this controversy. One
 factor is the presence of deep-seated intuitions that such a test is fundamentally
 incoherent. Perhaps the most compelling objection is that, given a
 set of discrete-time samples, one can construct an unaliased
 continuous-time series which exactly fits those samples. Therefore, the
 samples alone can not show that the original time series was aliased.
 The second factor prolonging the debate
 has been an inability of its proponents to unseat those objections.
 In fact, as is shown here, all objections can be met and the test as stated is
 correct. In particular, the role of stationarity as knowledge in
 addition to the sample values
 turns out to be crucial. Under certain conditions, including those
 addressed by the bispectral aliasing test, the continuous-time signals
 reconstructed from aliased samples are non-stationary.
 Therefore detecting aliasing in (at least some) stationary continuous-time processes both
 makes sense and can be done. The merits of the bispectral test for practical use
 are briefly addressed, but our primary concern here is its theoretical soundness.
\end{abstract}

\section{The Bispectral Aliasing Test}

The domain of the discrete-time bispectrum is the two dimensional
bifrequency $ \{\omega_1, \omega_2\} $ plane. Assuming a
real-valued discrete time series, the usual replication phenomenon
dictates that all non-redundant information is confined to the
square $ 0 \le \omega_1, \omega_2 \le \pi $. When one fully
accounts for symmetries, the non-redundant information in the
bispectrum is confined to a particular triangle inside this
square~\cite{harris, arl}.

This triangle naturally divides into two pieces. One piece is an
isosceles triangle and is unproblematic. The other piece, somewhat
unusual in shape, is the source of the controversy under
discussion. Naive consideration of this triangle shows that it
involves frequencies higher than the Nyquist frequency and
therefore must have something to do with aliasing. Hinich and
Wolinsky considered this more carefully and showed that the naive
intuition is correct: if the discrete time series arises from
sampling a stationary, band-limited, continuous-time process, and
if the sampling rate is sufficiently rapid to avoid aliasing, then
the discrete bispectrum is non-zero only in the isosceles
triangular subset of the fundamental domain. Conversely, if the
bispectrum of a sampled stationary continuous-time process is
non-zero in the outer triangle, then the sampling rate was too
slow to avoid aliasing.

It should be clearly understood that there is no assertion that
aliasing in general can be detected. The statement is not ``if a
signal is aliased, then the outer triangle will have a non-zero
bispectrum.'' Rather, the assertion is the converse, ``if the
outer-triangle shows a non-zero bispectrum, the (underlying)
continuous-time signal must have been aliased.''

At one level, this result is obvious and, in fact, the result was
initially so-regarded \cite{scharf}. However, doubt soon arose.
Perhaps the most important source for suspicion is the argument
based on reconstruction alluded to above.

In light of this objection, one is led to reconsider the
association of the outer triangle with aliasing. One can take the
position that there is no relation, as in \cite{frazer}. One can
decide that something is aliased, but that it is the bispectral
estimator rather than the signal. There is some plausibility to
this claim, for the frequencies that are involved in the outer
triangle are $\omega_1, \omega_2,$ and
 $\omega_1 + \omega_2 - 2\pi $. This seems to be the position of
Pflug et al. \cite{pflug}.

Or, one can try to delineate the conditions, if any, under which
the test makes sense. This was done by Hinich and Messer in
1995\cite{hm}. They confirmed the validity of the original
argument and stated its conclusions more carefully. In particular
they conclude that a non-zero bispectrum in the outer triangle
indicates a non-random signal or one of the following:
\begin{itemize}
  \item {a random, but non-stationary signal ;}
  \item {a random, stationary, but aliased signal, or;}
  \item {a random, stationary, properly-sampled signal which
  violates the mixing condition.}
\end{itemize}

We believe that the analysis of Hinich and Messer, while entirely
correct, did little to persuade the detractors of the test. In
particular their analysis did not address the reconstruction
objection and may have left the impression that the circumstances
for which the test applies are unlikely to be met in practice.

In this paper, we show that the reconstruction objection is far
from fatal. We further establish that stationarity is the only
property which is crucial to the test. Since this property is
required in order to define the bispectrum, one can legitimately
apply the aliasing test whenever one is entitled to compute a
bispectrum. Therefore the bispectral aliasing test is as
theoretically sound as the bispectrum itself.

\section{The Selection Rule and Brillinger's Formula}

The bispectrum, defined to be the triple Fourier transform of the
third-order autocorrelation, reduces to a function of two
frequencies since stationarity confines the spectrum to the plane
through the origin of the frequency domain perpendicular to the
vector (1,1,1).

\begin{equation}
\label{eq:selection}
        \mathfrak{F}_{123}(c_3(t_{1}, t_{2}, t_{3}))
          =b(\omega_{1},\omega_{2})
         \delta(\omega_{1}+\omega_{2}+\omega_{3})
\end{equation}

Another way of computing the bispectrum is to switch the order in
which one does the Fourier transforming and the ensemble
averaging. This leads to the following result.

\begin{equation}
\label{eq:fubini}
  b(\omega_{1},\omega_{2}) =  \langle X(\omega_{1}) X(\omega_{2}) X(\omega_3 = \omega_{1} + \omega_{2}) \rangle
\end{equation}

If the process is bandlimited and $X(\omega) = 0 $ for $|\omega| >
\pi$, then the bispectrum is confined to the intersection of the
$(1,1,1)$ plane and the $\pi$-cube (i.e. $(\omega_{1},\omega_{2},
\omega_{3}) \in (-\pi,\pi)\otimes(-\pi,\pi)\otimes(-\pi,\pi)$ ).
The plane and its projection onto the $(\omega_1,\omega_2)$ plane
is shown in Figure 1. Upon sampling with unit time step, one
obtains the usual replication in three dimensions.  (Doing
everything in 3-dimensions and projecting at the end keeps things
simpler and makes it easier to avoid errors.) In particular, one
gets that if the process is sampled at a frequency greater than
twice the highest frequency component, then the bispectrum is
confined to the replications of the tilted hexagon shown.

\begin{figure}[ht]
\centerline{ \epsfxsize 2.0in \epsffile{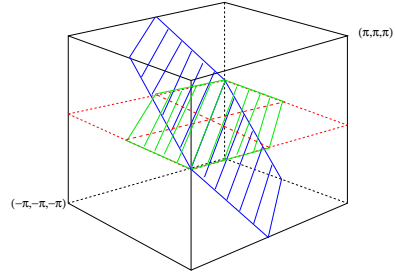}  }
\caption{ The origin of the bispectral fundamental domain.}
\label{fig:domain}
\end{figure}

The replication gives the discrete-time bispectrum $ b_d$:

\begin{equation}
\label{eq:replicate}
\begin{split}
   b_d(\lambda_1, \lambda_2, \lambda_3) =  \sum_{\omega_1 + \omega_2 + \omega_3 = 0}
      b(\omega_1, \omega_2, \omega_3).
\end{split}
\end{equation}

\noindent where $\omega_i = \lambda_i + 2\pi k $ for integer $k$.

Since the replication does not cause any overlaps, the outer
triangle remains empty.  This is the Hinich and Wolinsky aliasing
theorem. (Note that the outer triangle is equivalent to the bigger
triangle with vertices $(0,\pi,0),$ $(0,0,\pi)$ and $(0,\pi,\pi)$
by symmetries. See \cite{harris, arl} for details.)

\section{The Reconstruction Objection}

Suppose we have a stationary process $ x(t)$ and we undersample it
by sampling at $ t \in Z$. Then by convolving $ x(t)$ with the
appropriate $ sinc $ function we get a reconstructed process
$x_r(t)$. This new process will have exactly the same samples as
the original process and therefore exactly the same sampled
bispectrum: yet it is not aliased. Therefore for any process that
is undersampled, we have another process producing an identical
sampled process which is not undersampled, showing that that one
could not possibly detect aliasing via the bispectrum computed
from samples!

The rub here is the fact that the reconstructed signal will not
necessarily be stationary. Processes reconstructed from aliased
samples of continuous-time signals are generally cyclostationary
but not stationary. Some aliased processes do, in fact,
reconstruct into stationary processes. But in the class of
stationary signals for which the bispectral aliasing test gives
positive results, reconstruction from aliased samples produces
non-stationary processes.

To carefully illustrate this we will consider several stationary
processes generated by taking a periodic signal with period T and
giving it a random shift $\theta \in$ [0,T). First consider a
simple cosine process,

\begin{equation}
  \label{eq:cos}
 x(t) = \cos(\alpha \pi t + 2\pi\theta / T),
\end{equation}

\noindent where $\alpha$ = 1.5, T = 4/3, and $\theta$ is randomly
chosen from [0,4/3). Upon sampling and reconstruction we get the
cosine process given by

\begin{equation}
  \label{eq:cosrecon}
      x_r(t) = \cos(\hat{\alpha}\pi t + 2 \pi\theta/T),
\end{equation}

\noindent where $\hat{\alpha} = -0.5 $.  The key idea is that the
signal appears at the lower frequency as dictated by its
replication into the fundamental region of the frequency space,
(-$\pi,\pi$), but its reconstructed phase is the same as the
``source'' component phase.

Now consider

\begin{equation}
  \label{eq:twocos}
x(t) = \cos(\alpha \pi t + 2\pi\alpha\theta) +\cos(\beta \pi t +
2\pi\beta\theta),
\end{equation}

\noindent where $\alpha = 1.0$, $\beta = 3.0$, and $\theta$ is
chosen randomly from the interval $[0,\alpha^{-1})$. The
reconstructed process one gets is

\begin{equation}
  \label{eq:twocosrecon}
 x(t) = \cos(\alpha\pi t + 2\pi\alpha\theta) +\cos(\hat{\beta}\pi t +
2\pi\beta\theta),
\end{equation}

\noindent (where $\hat{\alpha}$ and $\hat{\beta}$ are the aliased
frequencies). Although the phase terms $2\pi\alpha\theta$ and
$2\pi\beta\theta$ are still uniformly distributed over $2\pi$ they
now correspond to different time shifts so we no longer have a
single shifted waveform. This process can easily be seen to be
cyclostationary, but not stationary. See Figure 2 \cite{vixie}.
Computation of the required expectations requires that one can
``average over the ensemble.'' Since this stationary process is
not ergodic, one can not get the result from a single realization
of the process. It is at this point that some differences in
perspective arise. Strictly speaking, in order to compute a
bispectrum one must perform the ensemble average. A single
realization does not suffice unless the process is ergodic.

\begin{figure}[ht]
\centerline{ \epsfxsize 2.0in \epsffile{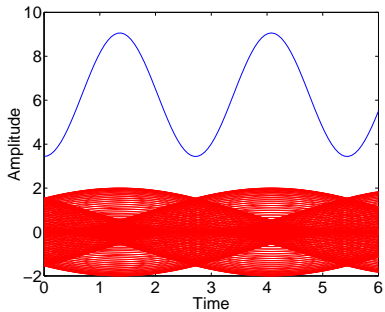}  } \caption{
Cyclostationarity of signal reconstructed from aliased samples.
The lower curve is the process envelope. The sampling interval is
$e$. The upper curve is the sixth moment, chosen for ease of
display.} \label{fig:cyclos}
\end{figure}

Finally, consider the process defined by

\begin{equation}
\begin{split}
  x(t) = & \cos((10/20)\pi t + 5\cdot2\pi\theta) +
        \\&  \cos((12/20)\pi t + 6\cdot2\pi\theta) +
        \\ & \cos((22/20)\pi t + 11\cdot2\pi\theta)
\end{split}
\end{equation}

\noindent where $\theta$ is chosen randomly from [0,1).

Because the phases of these components are in a fixed relation,
this process has a spike in the bispectrum at $\omega_1 =
10/20\pi$, $\omega_2 = 12/20\pi$, i.e., in the outer triangle. The
reconstructed signal is given by

\begin{equation}
\begin{split}
   x(t) = & \cos((10/20)\pi t + 5\cdot2\pi\theta)  +
   \\ & \cos((12/20)\pi t + 6\cdot2\pi\theta) +
   \\ & \cos(-(18/20)\pi t + 11\cdot2\pi\theta)
\end{split}
\end{equation}

\noindent which is not stationary. Therefore we have a signal with
nonempty outer triangle whose reconstruction is not stationary.
This situation is exactly what the bispectral test implies happens
whenever the outer triangle is nonempty. The loss of stationarity
causes the Fourier transform of the triple autocorrelation to
``move off'' of the (1,1,1) plane.

Therefore, if one knows (or is willing to assume) that the process
which generated the observed samples was stationary, one can rule
out the unaliased reconstruction as the source of the samples. In
a sense, the continuous time signal reconstructed from aliased
samples of an original time series is a ``measure zero'' object.
This result is very surprising to most people's intuitions. It is
studied further in Vixie, Sigeti and Wolinsky \cite{vixie}.

\section{The Replication objection}

Upon looking at Equation \ref{eq:fubini} one may observe that even
if X($\omega$) = 0 for $|\omega| > \pi$, sampling effectively
fills in the spectrum at higher frequencies. This is the basis for
the objection appearing in Swami \cite{swami-1993}. This concern
is addressed as follows. While the spectrum does indeed fill out
upon sampling, the undesired expectations remain zero. Consider a
(statistically stationary) ensemble constructed by uniformly
translating a periodic or finite-duration waveform $x(t)$. Two
operations are necessary to produce the discrete-time ensemble; a
uniform shift in time over a period $T$, which introduces linear
phase factors, and sampling, which produces spectrum replication.
These operations do not commute: i.e., one wants to time-shift the
waveform first and then sample, rather than to shift its samples.
For the shifted samples $x_s(t + \theta)$ one finds

\begin{equation}
   \mathfrak{F}(x_s (t + \theta)) = e^{i\theta\omega} \mathfrak{F}(x_s(t))
\end{equation}

But for the sampled shifted waveforms $x(t+\theta) {\mid}_s$,
i.e., the waveforms needed to construct a stationary ensemble, the
phase of the original signal is propagated to higher frequencies
periodically rather than linearly. This difference leads to the
vanishing of unwanted expectations.

For example, consider the process given by the randomly shifted
sum of unit amplitude cosine waves with frequencies at $n/20$
(rad/s) where $n$ takes integer values from $1$ to $19$. The
sampled spectrum has components at $\omega_1 = 10\pi/20,$
$\omega_2 = 11\pi/20$ and $ \omega_3 =-21\pi/20$ but the average

\begin{equation}
  \langle X(\omega_1)X(\omega_2)X(\omega_3) \rangle
\end{equation}

\noindent reduces to

\begin{equation}
  \langle
  e^{i10\theta\pi}e^{i11\theta\pi}e^{i19\theta\pi}\rangle_{\theta}
\end{equation}

\noindent where $\theta$ is chosen with uniform probability from
[0,1). This average vanishes. Therefore, the potential
contribution in the outer triangle is zero because averaging kills
it. This is in contrast to the case where the average is zero
because the spectral amplitudes are themselves zero (as in the
proof of the aliasing test).

\section{Empirical counter-examples}

Other objections to the test have been made. Frequently these
objections involve a (purported) counter-example to the bispectral
aliasing test. A particularly clear example is provided by Frazer,
Reilly and Boashash \cite{frazer}. Here the authors do two things.
They present an example of an aliased signal which the aliasing
test fails to mark as aliased. The example is unproblematic:
neither the aliasing test nor any aliasing test we are aware of
will detect all aliased signals. It is not, however, a
counter-example to the test. Since there is nothing in the outer
triangle, the bispectral aliasing test makes no assertion
regarding the presence of aliasing.

The other example the authors provide is more interesting. It
consists of a signal involving coupled sinusoids at $ \omega_1 =
0.3125 Hz$, $ \omega_2 = 0.25 Hz $ and $\omega_3 = .4375 Hz$ and
the authors show that there is a peak in the outer triangle under
conditions which rule out aliasing. As the authors note these
frequencies sum to 1 Hz ((the sampling rate). Under these
conditions the authors are correct in asserting that the aliasing
test gives a positive result, which they believe to be incorrect.
However, what the aliasing test actually indicates is that this
signal is non-stationary. The particular interaction which the
authors have constructed is not one for which the continuous-time
selection criteria is met, i.e., the frequencies involved do not
sum to zero. Samples of this signal do meet the discrete-time
stationarity condition and so a non-zero bispectrum is possible in
the outer triangle.

One can look at these results in various ways. Our position is
that neither example constitutes a counter-example to the validity
of the aliasing test in theory, though they both show that the
test is limited in practice. The first example shows that there
are aliased signals which the test does not see. This is obvious
anyway since there are signals with zero bispectrum whose samples
can be aliased. The second example shows that the term ``aliasing
test'' must be restricted to stationary signals. As stated
earlier, this restriction is inherent in the definition of the
bispectrum.

\section{Conclusions}

So, is this something for nothing?  How can one get information
about higher frequency amplitudes from what is usually thought of
as Nyquist-limited data? The answer is of course that the
assumption of stationarity is far from nothing. But, to exploit
stationarity one must be able to perform the ensemble averaging
indicated in the definition of the bispectrum. This implies that
one must either have an ergodic process or have access to
sufficiently many sample paths.

It is certainly possible that, in practice, the bispectrum can be
usefully applied to signals for which there is no theoretical
justification. For such uses the aliasing test is silent. However,
it is essential that a clear understanding of the fundamental
properties of higher-order spectra be available. And the present
authors believe that correct understanding of the outer triangle
leads to deeper insight of the meaning of the bispectrum in
general.

\bibliographystyle{plain}
\bibliography{the_bib}

\end{document}